# Electrospun amplified fiber optics

*Giovanni Morello[a,b], Andrea Camposeo[a,b], Maria Moffa[a], and Dario Pisignano[a,b,c],\**

[a] Istituto Nanoscienze-CNR, via Arnesano, I-73100 Lecce, Italy.

[b] Center for Biomolecular Nanotechnologies @UNILE, Istituto Italiano di Tecnologia, Via Barsanti, I-73010 Arnesano (LE), Italy

[c] Dipartimento di Matematica e Fisica "Ennio De Giorgi", Università del Salento, via Arnesano I-73100 Lecce, Italy

*Corresponding author: dario.pisignano@unisalento.it

ABSTRACT

A lot of research is focused on all-optical signal processing, aiming to obtain effective alternatives to existing data transmission platforms. Amplification of light in fiber optics, such as in Erbium-doped fiber amplifiers, is especially important for an efficient signal transmission. However, the complex fabrication methods, involving high-temperature processes performed in highly pure environment, slow down the fabrication and make amplified components expensive with respect to an ideal, high-throughput and room temperature production. Here, we report on near infrared polymer fiber amplifiers, working over a band of about 20 nm. The fibers are cheap, spun with a process entirely carried out at room temperature, and show amplified spontaneous emission with good gain coefficients as well as low optical losses (a few cm$^{-1}$). The amplification process is favoured by the high fiber quality and low self-absorption. The found performance metrics promise to be suitable for short-distance operation, and the large variety of commercially-available doping dyes might allow for effective multi-wavelength operation by electrospun amplified fiber optics.

## INTRODUCTION

In the information era, the capability to transmit data in an efficient way represents the touchstone for establishing technological leadership.[1,2,3] In particular, yielding efficient transmission of optical data with minimal attenuation is a traditional challenge. Nowadays, the chosen method is based on amplifiers able to counterbalance the intrinsic attenuation of transmission channels. While silica fiber amplifiers doped by Erbium or other rare-earth ions[4-8] are generally used for long-distance operation, polymer fibers exhibiting optical gain can be advantageous for applications relying on short-range signal transmission with a large number of nodes, including domotics and biomedical networks. The operation of such fibers has been largely limited to the visible range hitherto.[9-12] In these components, the amplification process is promoted by the stimulated emission occurring in chromophores or active dopants once the population inversion is reached and a spectrally-matching signal passes through the fibers. In this respect, an optical amplifier can be considered as a lasing system lacking feedback.

In a typical working architecture, one end of a fiber or of a waveguide is doubly coupled, namely it is interfaced to both the transmission line and to the pumping source, the latter leading to the population inversion.[9-12] A number of processes, such as self-absorption by the active materials and out-coupled light at the fiber surface contribute to optical losses, which limit the system efficiency. In addition, spurious spontaneous emission is a common source of noise since so produced light does not have coherence characteristics as required for amplification. Moreover spontaneous emission could be in turn amplified once the population inversion is reached, thus competing with the main signal.

In this framework, most of the used silica-based systems present a drawback in the complexity of their fabrication, which may involve vapour deposition techniques or quite complex chemical methods, interfacial gel polymerization, preforming, etc.. Such processes may require a highly pure environment, and may be slow or generally expensive. Polymeric systems, on the other hand, can be valuable alternatives for many applications in industrial automation and multimedia connections in cars.[13,14] Plastic materials take advantage from the wide versatility of available doping compounds and molecular dyes, making them appealing for the development of all-optical amplification schemes.[15] In addition, one should mention that coupling miniaturized fibers with conventional fiber optics is often an issue.[2] For instance, efficient evanescent coupling with tapered optical fibers has been recently used to this aim to successfully couple a variety of glass, polymer and metal miniaturized fibers.[16,17] For all these reasons, the achievement of alternative, low-cost and versatile amplified fibers, allowing different experimental configurations to be developed, is highly desired.

Over the past two decades, organic gain media have attracted great interest as light-emitting sources, lasers and optical amplifiers.[2,18] Drawn and spun active nanofibers[3,19] have recently emerged as building blocks for photonics, exhibiting all the advantages of organics, including the low cost of materials and processing, the simplicity of their fabrication methods, as well as the possibility to engineer their optical properties, photoluminescence quantum yield and Stokes shift.[20] These fibers are also fully compatible with dyes,[9,15] rare-earth ions[21,22] and nanocrystal dopants.[22,23,24] Furthermore, fibrous organic amplifiers may show large stimulated emission cross-section, which makes them attractive for optical sensing, on-chip spectroscopy, data communications and processing,[18] including light amplification and logic operation. In particular, electrospinning is a low-cost, high throughput and room-temperature process to make fibers.[25] Hence, electrospun fiber optics would present clear advantages, including low power consumption for their production as well as for their operation due to the low thresholds for optical gain and high fluorescence efficiency.[20]

Here, we report on near infrared (NIR) amplifiers based on individual, electrospun plastic fibers with length up to many mm, and on their arrays. We focus on molecular systems emitting at wavelengths around the first transmission window of conventional optical fibers. Furthermore, the NIR region (wavelengths ≤ 950 nm) is also of interest for many biomedical applications such as in-vivo fluorescence imaging and sensing, due to the low auto-fluorescence from biological tissues and deep penetration under human skin.[26-28] The here proposed system shows an amplification of 14 dB with a length of 8 mm, and works on a wavelength band of 20 nm around the peak at 950 nm, with good optical gain and losses metrics compared to other micro-fabricated polymer fibers.[9-12]

**RESULTS AND DISCUSSION**

We electrospin poly(methylmetacrylate) (PMMA) doped with 2-[2-[3-[[1,3-dihydro-1,1-dimethyl-3-(3-sulfopropyl)-2H-benz[e]indol-2-ylidene]ethylidene]-2-[4-(ethoxycarbonyl)-1-piperazinyl]-1-cyclopenten-1-yl]ethenyl]-1,1-dimethyl-3-(3-sulforpropyl)-1H-benz[e]indolium hydroxide, inner salt, compound with $n,n$-diethylethanamine(1:1) (hereinafter referred to as IR 144). The basic spectral features of the emitting system are shown in Figure S1 in the Supporting Information, highlighting a 320 meV gap from the absorption maximum to the peak of the amplified spontaneous emission (ASE). Uniaxially aligned and randomly oriented fibers can be easily obtained by varying the geometry of the collecting surfaces in the electrospinning set-up. Examples of aligned fibers are displayed in Figures 1a,b. The fibers may exhibit a ribbon shape, resulting from a rapid solvent evaporation and consequent solidification of the jet surface, followed by a collapse of the so-formed sheath.[29] Such fibers feature cross-sectional dimensions which can range from the scale of 1 μm to 500 μm × 70 μm (Figure S2a) depending on the process parameters. The IR 144 dye is uniformly distributed in the host matrix (Figure S2b). Figure 1c sketches the waveguiding properties of our fibers, measured by micro-photoluminescence (μ-PL) experiments. By exciting the fiber at a varying distance, $D$, from the exit termination, we observe some luminescence signal escaping the fiber body (as in the right spots in Figure 1c), and an emission guided along the fiber longitudinal axis. The out-coupled signal from the fiber termination decreases by increasing the distance $D$, as highlighted by the vertical arrows in Figure 1c. Optical losses can be ascribed to a number of physical mechanisms, such as residual self-absorption.[15] The analysis of the emitted intensity ($I_{PL}$) as a function of the tip-excitation distance ($D$) gives insight into the total losses ($\gamma$) in the fibers by fitting the experimental data by the expression $I_{PL} = I_0 \cdot e^{-\gamma D}$ (Figure 1d). We find $\gamma = (226 \pm 10)$ cm$^{-1}$ for fibers under continuous wave excitation, a value in line with or outperforming electrospun polymer fibers with emission in the visible spectral range.[30]

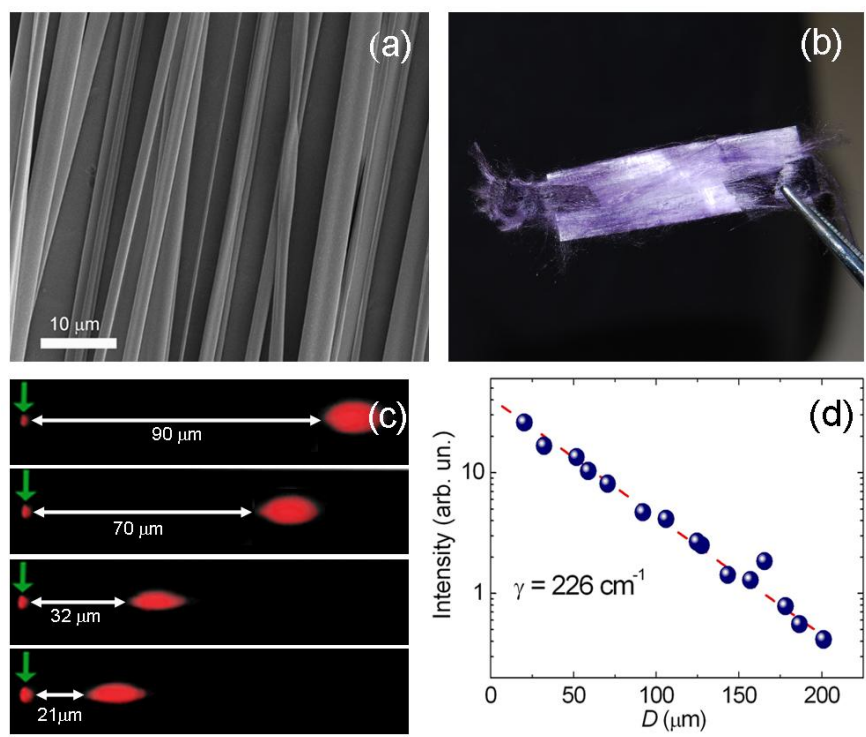

**Figure 1.** (a) Scanning electron micrographs of fibers doped with IR 144 and aligned in a bundle. Each individual fiber appears uniform along its length and free of defects. (b) Photograph of a free-standing bundle of uniaxially aligned IR 144-based fibers. (c) μ-PL micrographs of IR 144-based fibers, pumped by a laser excitation beam tightly focused in spots (right bright spots in each micrograph) at variable distances, $D$, from the fiber termination (left, smaller spots). This termination is highlighted by vertical arrows, and corresponds to the point from which the out-coupled PL signal is collected, after being transmitted along the fiber longitudinal axis. (d) Corresponding PL intensity vs. $D$ (symbols) and best fit to equation, $I_{PL} = I_0 \cdot e^{-\gamma D}$ (dashed line).

We then investigate the gain properties both in bundles and in single fibers, and the amplification of transmitted optical signals, by ASE experiments. Figure 2a displays the fiber emission spectra, collected at different excitation fluences and pumping by a stripe along the alignment direction in the bundle.

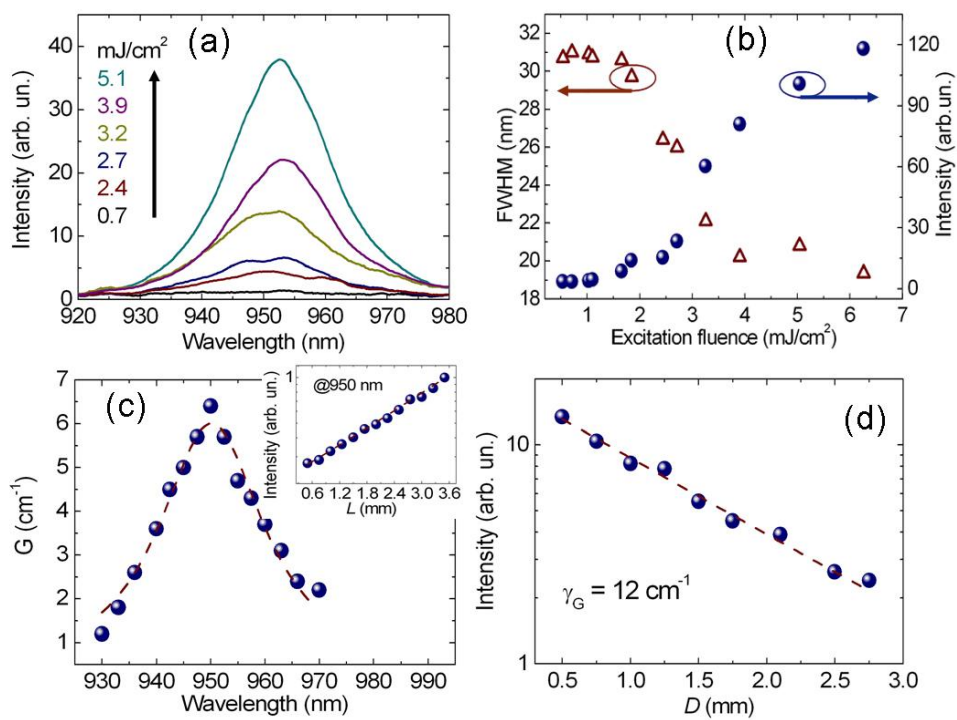

**Figure 2.** (a) ASE spectra from fibers for various excitation fluences. (b) ASE intensity (full symbols) and FWHM (empty symbols) vs. excitation fluence. (c) Wavelength dependence of the net gain of fibers. The dashed line is a guide for the eyes. Inset: net gain analysis performed at 950 nm. (d) Analysis of losses under pulsed excitation. Experimental values of the ASE intensity (symbols) are plotted *vs.* the distance of the exciting stripe from the emitting edge of the substrate. The dashed fitting line is obtained by the equation: $I_{PL} = I_0 \cdot e^{-\gamma_G D}$. In (c) and (d) the excitation fluence is 3.3 mJ/cm$^2$.

The emission is peaked at about 950 nm (Figure 2a), with minor spectral fluctuations as typical of ASE. In particular, intensity undergoes a rapid increase when optical pumping exceeds a value around 1.5 mJ/cm$^2$, then following a super-linear dependence (in the range 2-4 mJ/cm$^2$) until saturation is approached (at about 5-6 mJ/cm$^2$, Figure 2b). Spontaneous emission keeps low (Figure S3) and the full width at half maximum (FWHM) of the emission spectra decreases as a consequence of the amplification, reaching a value as low as 19 nm. ASE is a fundamentally thresholdless process,[31] however an experimental threshold can be conventionally defined as the value of the excitation fluence at which the measured FWHM reaches the average of the FWHM of the spontaneous emission and of that of ASE.[32] This approach provides a threshold excitation fluence of about 2.5 mJ/cm$^2$ for IR 144-doped electrospun fibers. At fluences of 10-12 mJ/cm$^2$ photobleaching effects become significant and a decrease of ASE is observed (Figure S4), whereas evidence of physical damage (laser ablation) in fibers is found at fluence exceeding 40 mJ/cm$^2$. This class of ASE-showing spun fibers opens interesting perspectives for signal transmission and amplification. In particular, if the transmitted signal wavelength well matches that of the maximum material gain, the amplification process could proceed in an efficient way. The efficiency basically depends on the pumping fluence and the signal level,[33,34] and is decreased by possible detrimental effects such as noise and/or losses depleting excited states. Fibers based on four-level active molecular systems, for which the gain threshold and self-absorption are minimized (due to large Stokes shifts),[2,9,15] would be advantageous since showing minimal noisy emission.[2, 5, 9, 15, 18, 31] We determine the wavelength dependence of the net gain, $G(\lambda)$, along our fibers by fitting the data of PL intensity ($I_L$) vs. excitation stripe length ($L$) at each wavelength, to the equation:

$$I_L = \frac{I_p A(\lambda)}{G(\lambda)} \cdot \left[ e^{G(\lambda) \cdot L} - 1 \right] \qquad (1)$$

where $I_p$ represents the pump intensity and $A(\lambda)$ is a factor accounting for the spontaneous emission cross-section. Results are shown in Figure 2c and evidence a maximum $G$ value of almost 7 cm$^{-1}$ at a wavelength of 950 nm.

Optical losses under pulsed pumping conditions are determined by moving an excitation stripe of fixed length (4 mm), away from the emitting termination of fibers, thus increasing the separation distance, $D$, namely the length of the unexcited region. This experiment leads to measure a loss coefficient, $\gamma_G = (12 \pm 1)$ cm$^{-1}$ (Figure 2d), a value much smaller than that obtained for the waveguided spontaneous emission (Fig. 1d). This difference can be rationalized by considering that ASE (peaked at 950 nm) is affected by lower self-absorption losses compared to spontaneous emission (peaked at about 865 nm). Such different behavior can be taken into account by calculating the ratio of the absorption coefficients measured at the absorption ($\alpha_{abs}$) and emission ($\alpha_{PL, ASE}$) peaks:[35] $S_{PL,ASE} = \alpha_{abs} / \alpha_{PL, ASE}$, where the PL and ASE subscripts indicate the spontaneous emission and ASE peak wavelength, respectively. Higher $S$ values correspond to a lower contribution of self-absorption to propagation losses. In our system we find $S_{PL} = 3$ and $S_{ASE} = 35$, whose ratio (about 12) is in good agreement with the measured ratio of loss coefficients, $\gamma / \gamma_G \cong 19$. The here measured $\gamma_G$ is slightly higher than that found in visible-emitting conjugated polymer slabs made of poly(9,9-dioctylfluorene) or blends of poly(p-phenylene vinylene) derivatives.[36,37] Given the reduced size of electrospun fibers and the relatively lower refractive index of PMMA compared to conjugated polymers (both features in principle increase optical losses from waveguides), this result supports the high surface quality, the low density of light-scattering defects and the reduced self-absorption of the here micro-fabricated amplifiers.

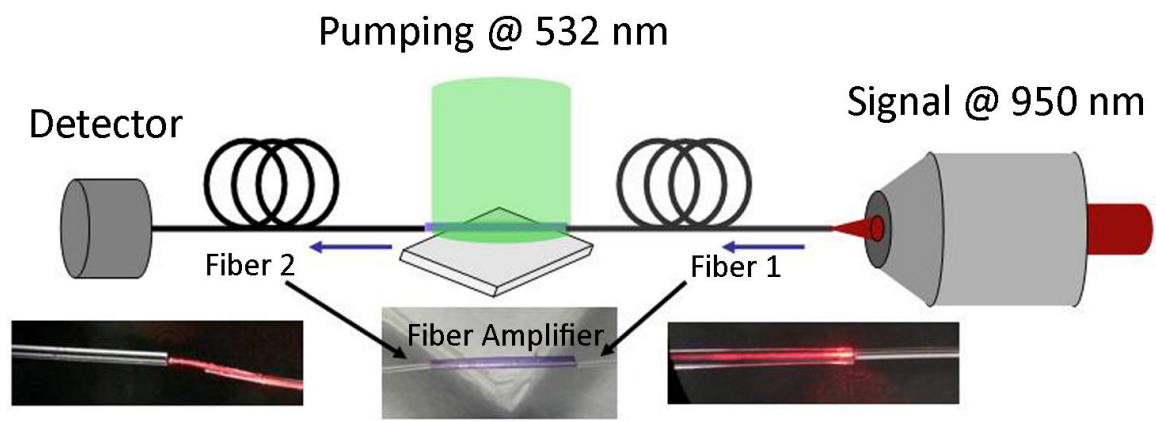

**Figure 3.** Experimental configuration of the electrospun fiber amplifier. The probe signal (at 950 nm, 960 nm, or 970 nm) is launched into a single mode fiber (Fiber 1, core diameter of 6.6 μm) and passes through a single electrospun fiber, deposited on a PDMS substrate and providing amplification. The signal is then coupled to a second optical fiber (Fiber 2) and collected by a NIR detector. Pumping is performed by a collimated laser beam ($\lambda$ = 532 nm). Photographs of a representative single electrospun fiber butt-coupled to Fiber 1 and 2 are also shown in the bottom insets.

A simple proof-of-principle experiment can be based on a conservative approach in terms of amplifier working conditions, with the aim to minimize noisy signals and avoid saturation effects and without the need for pump-signal synchronization. We chose a non-monochromatic signal, with central wavelength corresponding to the maximum optical gain (at 950 nm) and a linewidth comparable to that of the fiber gain. The pumping fluence is chosen in the large range of superlinear variation of ASE intensity. Fibers are excited by a wavelength of 532 nm through a side-pump scheme based on a collimated, uniform pump beam (Figure 3).[38] Such a scheme substantially differs from configurations which are typically used in rare-earth-doped fiber lasers, such as the end-pump scheme and the more recent distributed side-coupled cladding-pumping scheme, which has been designed to avoid the high temperature rise occurring at the terminations of end-pumped, high-power fiber lasers.[39] The choice of a side pump scheme is here due to the high absorption coefficient of organic nanofibers compared to rare-earth fiber laser. Here the dye absorption at the pumping wavelength ($\lambda = 532$ nm) is of the order of $10^6$ dB/m, which is much higher than values in rare earth-doped fiber lasers at their pumping wavelengths (1-10$^2$ dB/m).[40,41] This high absorption would lead to a substantial attenuation of the pumping beam within a depth of about 20 μm in the plastic medium, which makes side-pump schemes more effective for electrospun fiber amplifiers. The single, active electrospun fiber is then deposited on a polydimethylsiloxane (PDMS) substrate and coupled to an input and an output mono-modal fiber in butt-coupling configuration, as shown in Figure 3. Figure 4 shows the resulting differential transmission, $\Delta T/T$, measured in a single fiber at different pump fluences (Figure 4a) and for different wavelengths (Figure 4b). For each data point, we record the intensity of four transmitted signals ($I_{00}$, $I_{01}$, $I_{10}$, and $I_{11}$). $I_{00}$ corresponds to a configuration with both the pump and the input signal turned-off, $I_{01}$ to that with pump off and signal on, $I_{10}$ to that with pump on and signal off, and $I_{11}$ represents the transmitted intensity with both the input signal and the pump on, respectively. The relative increase in the intensity of the transmitted signal, $T$, is then obtained by the ratio $\left(I_{11} - I_{10}\right)/\left(I_{01} - I_{00}\right)$,[34] since $I_{11} - I_{10}$ is the transmitted signal intensity ($T_F$) at a given optical excitation density (i.e. at a given pump fluence), whereas $I_{01} - I_{00}$ represents the transmitted signal ($T_0$) in absence of optical pumping. The differential transmission, $\Delta T$, is consequently estimated as:

$$\frac{\Delta T}{T} = \frac{T_F - T_0}{T_0}.$$ (2)

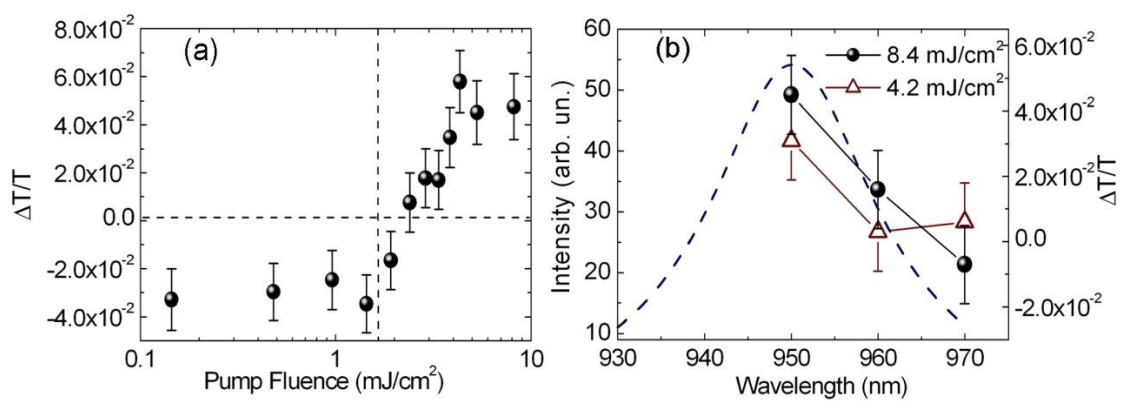

**Figure 4.** (a) Plot of the differential transmission along the electrospun fiber amplifier (at 950 nm) as a function of the excitation fluence. A net amplification of the transmitted signal is found at fluences larger than about 2 mJ/cm$^2$. (b) Spectral dependence of the transmission at two distinct excitation fluences (full circles and empty triangles, right vertical scale) and gain spectrum (dashed line, left vertical scale). The maximum amplification is reached in correspondence to the maximum optical gain (Fig. 2c).

Three regimes could be identified, as featured by different transmission trends. Up to pumping fluences of about 1-2 mJ/cm$^2$, the signal undergoes attenuation up to 4%, which can be related to photoinduced absorption phenomena leading to a weaker transmission upon pumping. In the range of fluence from 2 to 4 mJ/cm$^2$ the output signal increases with respect to the input, which is ascribable to the stimulated emission from IR 144 occurring at the signal wavelength (950 nm in Figure 4a). At fluence above 4 mJ/cm$^2$, the differential transmission keeps unchanged, a fingerprint of the saturation of gain for which the population inversion cannot be further enhanced, or slightly decreases analogously to the behaviour found for ASE (Figure 2b). As a further confirmation of the effective amplification process, we analyse the signal transmission at 960 nm and at 970 nm (Figure 4b), finding a lower gain in agreement with the spectrum obtained by the ASE characterization (Figure 2c). Importantly, the amplification of about 14 dB on a length of 8 mm, is comparable or larger than the typical maximum gain obtained in other dye-doped fiber amplifiers[12,38] as well as in rare-earth-doped fibers.[42] Considering their small size, these new fibrous systems could be considered as promising media for short-distance as well as pulsed amplification uses.

## CONCLUSIONS

In conclusion, we demonstrate amplified fiber optics realized by electrospinning. Doped with NIR laser dyes, these plastic fibers allow optical amplification to be observed with good performances, presenting a regime of operation of 20 nm, flexibility and lightweight, capability of interfacing with conventional optical fibers, and unique fabrication advantages. These components are promising candidates for the future development of all-plastic, micro- to macroscale optical platforms for signal transmission and processing.

**MATERIALS AND METHODS**

*Fibers fabrication and morphological characterization*. Bundles of uniaxially aligned fibers are fabricated by electrospinning a PMMA-IR 144 solution onto a rotating disk collector (diameter = 8 cm, thickness = 1 cm, 4000 rpm). Briefly, the solution is prepared by dissolving PMMA (375-650 mg/mL, depending on the desired final fiber thickness) with IR 144 (3.1 mg/mL) in chloroform, and sonicated at 40°C for 6 hours. The solution is loaded into a 1 mL syringe and delivered at a constant flow rate (1 mL/h) through a metal needle (21 gauge) connected to a high-voltage power supply (EL60R0.6–22, Glassman High Voltage, High Bridge, NJ). Upon applying voltage (10 kV), a polymer solution jet is ejected from the needle and fibers are deposited on square ($1\times1$ cm$^2$) quartz substrates mounted at the edge of the rotating disk. The morphology of the fibers is inspected using scanning electron microscopy (SEM, FEI Company, Hillsbora, Oregon-USA) following thermal deposition of 5 nm of Cr. The average diameter of fibers is ($1.8\pm0.8$) μm, calculated from SEM micrographs by an imaging software (WSxM, Nanotec Electronica, Madrid, Spain) and analyzing a total number of at least 100 fibers.

*Confocal and waveguiding measurements*. Confocal micrographs (Olympus FV-1000) are collected by laser scanning ($\lambda_{exc}$ = 405 nm) a field of fibers, in epilayer configuration. The laser beam passes through an objective lens (40× and numerical aperture, NA = 0.75), thus impinging onto the fibers and exciting fluorescence. The PL signal is then collected through the same microscope objective and analyzed by a multianode photomultiplier. A comprehensive description of waveguiding measurements by μ-PL is reported in Ref. 30. Briefly, the fibers are deposited on a quartz substrate and positioned in an inverted microscope. The fibers are cut, thus protruding from the substrate by about 1 mm and being suspended at one end. A laser beam ($\lambda_{exc}$ = 408 nm) passes through an objective lens (20×, NA = 0.5) and excites the fibers, whose emission is then collected by the same objective, dispersed by a 0.33 m long monochromator and detected by a charged coupled device (CCD). Alternatively, the fiber emission can be directed to another CCD camera (Leica, DFC 490) without spectral dispersion, for imaging. The fibers are side-pumped at distance, $D$, from the freestanding tip. By varying the $D$ value, the PL intensity is measured vs. the distance, normalizing data to the PL intensity collected from the excitation spot in order to account for local, minor sample disuniformities.

*ASE characterization.* The fibers are put under vacuum and excited by the third harmonic of a pulsed Nd:YAG laser ($\lambda_{exc}$ = 355 nm, repetition rate = 10 Hz, pulse duration = 10 ns). The excitation spot is focused on the samples in a stripe (maximum length = 4 mm) and the signal collected from one edge of the substrate is measured with a monochromator and a CCD. For net gain characterization the stripe length is varied by a controllable slit, keeping fixed the excitation fluence. Optical losses under pulsed excitation conditions are investigated by varying the distance of the stripe from the fiber edge, at fixed fluence.

*Transmission measurements.* As input signal, the light from a halogen lamp is dispersed by a grating (600 lines/mm, blazed at 1000 nm) in order to have an almost monochromatic beam with FWHM of about 10 nm. To avoid saturation effects, the signal power density is kept low (a few W/cm$^2$). The signal is passed through an objective lens (4×, NA = 0.16) and coupled to a single-mode optical fiber (core diameter = 6.6 μm, NA = 0.13). For the transmission measurements a single electrospun fiber is butt-coupled to the input optical fiber at one end and to an identical optical fiber (output fiber) to the opposite end. The signal passing through the system is analyzed by a NIR, nitrogen-cooled CCD. The pump is provided by the second harmonic of a pulsed Nd:YAG laser (532 nm, repetition rate = 10 Hz,  pulse duration = 10 ns), shaped in a uniform beam along the body of the electrospun amplifier. Measurements are performed by varying the pump fluence over more than one order of magnitude, for values corresponding to those relevant to the stimulated emission dynamics.

ACKNOWLEDGMENT

The authors want to thank Dr. P. Del Carro for help in the morphological characterization. The research leading to these results has received funding from the European Research Council under the European Union's Seventh Framework Programme (FP/2007-2013)/ERC Grant Agreement n. 306357 ("NANO-JETS").

# Supporting Information

# Electrospun amplified fiber optics

*Giovanni Morello[a,b] Andrea Camposeo[a,b]Maria Moffa[a], and Dario Pisignano[a,b,c,*]*

[a] Istituto Nanoscienze-CNR, via Arnesano, I-73100 Lecce, Italy.

[b] Center for Biomolecular Nanotechnologies @UNILE, Istituto Italiano di Tecnologia, Via Barsanti, I-73010 Arnesano (LE), Italy

[c] Dipartimento di Matematica e Fisica "Ennio De Giorgi", Università del Salento, via Arnesano I-73100 Lecce, Italy

*Corresponding author: dario.pisignano@unisalento.it

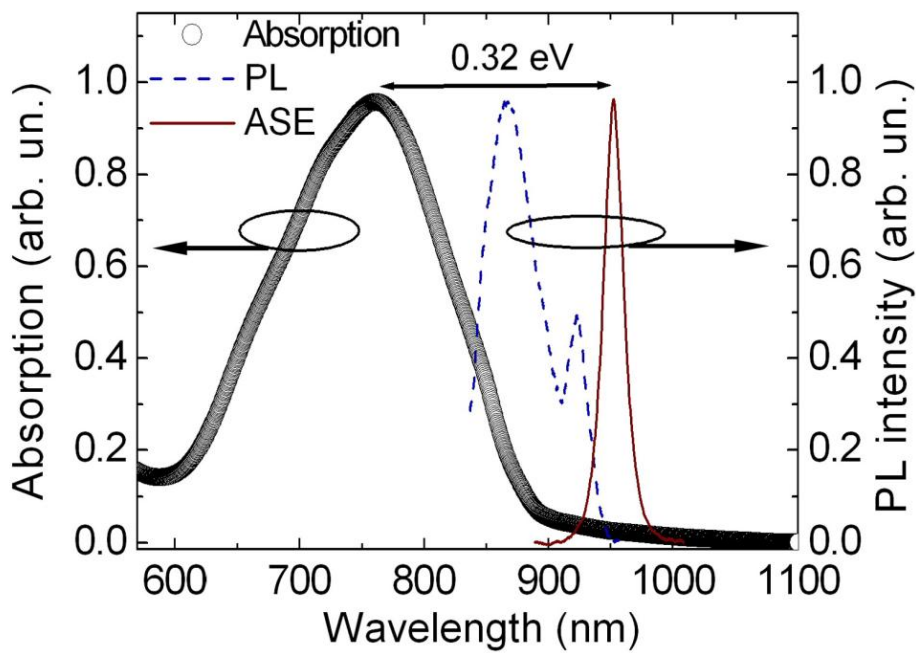

**Figure S1.** Normalized absorption, photoluminescence and ASE spectra of IR144-based films.

The presence of a double emission band is observed in the photoluminescence spectrum.

Absorption-ASE shift = 0.32 eV.

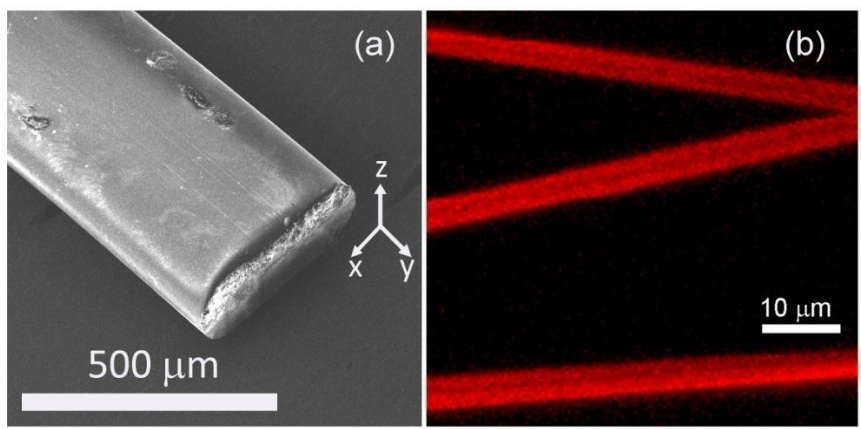

**Figure S2.** (a) SEM micrograph of a thick electrospun fiber (cross-sectional size = 470×70 μm²).

(b) Confocal micrograph of a few micrometer-sized, IR 144-doped fibers. The fibers show

uniform emission along their main axis, without evidence for microscale defects or significant dye clustering.

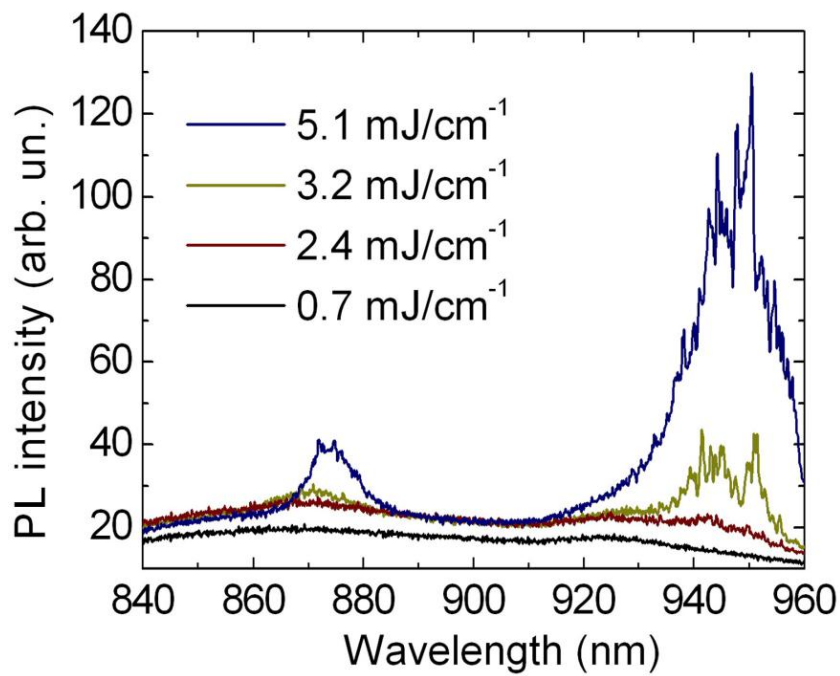

**Figure S3.** Full range emission of IR 144-doped films vs. excitation fluence, acquired by a Si CCD. Two spectral bands show ASE, with different threshold and peak features. The 870-880 nm peak has a threshold at about 3 mJ/cm$^2$. The 950 nm peak shows higher intensity together with very sharp features, ascribable to random feedback due to light scattering on micrometric clusters, whereas the apparent low signal at higher wavelengths is due to low detector sensitivity. The random lasing is largely removed in electrospun fibers, where the fabrication conditions contribute to well distribute the dye in the PMMA matrix.

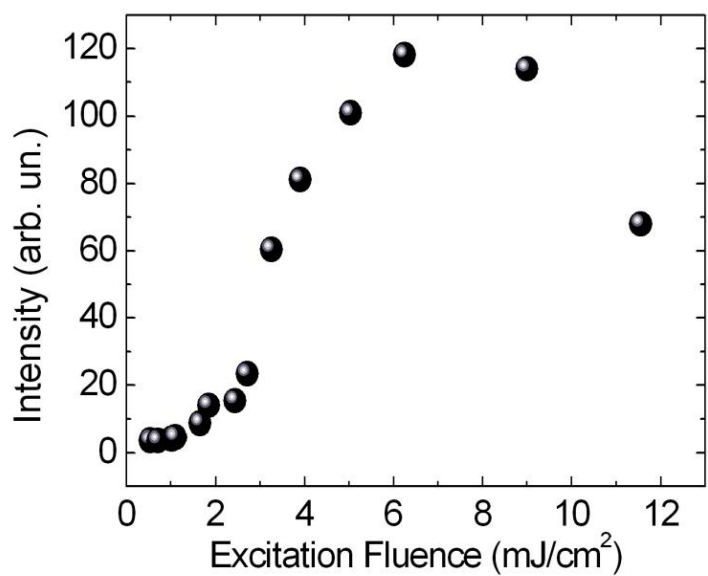

**Figure S4**. ASE intensity of IR 144-doped fibers, for excitation fluences up to about 12 mJ/cm$^2$.

Photobleaching is seen at 10-12 mJ/cm$^2$.